# Capacitively Coupled GaAs p-i-n/Substrate Photodetector with Ohmic Contacts on Lightly Doped n-GaAs for Hard X-Ray Imaging

V.G. Harutyunyan[1,2*], S.D. Zilio[2], M. Colja[3], M. Cautero[3], G. Cautero[3], L. Sbuelz[2], D. Curcio[2], G. Biasiol[2]

[1]*Institute of Applied Problems of Physics, National Academy of Sciences of the Republic of Armenia, 25 Hrachya Nersisyan, Yerevan 0014, Armenia,*
[2] *CNR-Consiglio Nazionale delle Ricerche, Istituto officina dei materiali, Area Science Park, Strada Statale 14 km 163.5, Basovizza, Trieste 34149, Italy*
[3]*Elettra-Sincrotrone Trieste S.C.p.A., Area Science Park, Strada Statale 14 km 163.5, Basovizza, Trieste 34149, Italy*

## Abstract

In this work, we present a capacitively coupled GaAs $p^+$-i-n/substrate photodetector (CC-GaAs PIN/S PD), which also represents a preliminary step toward 3D detection (x, y, time) of high-energy X-ray pulses. Although the final 3D detector will be based on a separate absorption and multiplication avalanche photodiode (SAM APD) design, the present device exhibits characteristics that offer valuable insights into the performance expected once a multiplication layer is incorporated into the final device.

In particular, we present a fabrication strategy that employs multi-step annealing in the low-temperature range of 280–330 °C to achieve Cr/Au ohmic contacts on lightly doped n-GaAs, which is also required for the photodetector architecture. Simultaneously, the same contact preparation process was applied to $p^+$-GaAs. Furthermore, the fabricated CC-GaAs PIN/S PD includes an additional contact designed to reduce leakage current by applying the same bias as that of the anode. Measurements performed using an 80 MHz laser demonstrated the photodetector's ability to detect pulses corresponding to ~$10^6$ electrons per pulse.

[*]**E – Mail:** *vgharutyunyan@gmail.com*

# 1. Introduction

Hard X-ray detection has gained significant attention for its wide range of applications, including material science (Falch et al., 2017), nondestructive testing, medical imaging, and flash X-ray radiography (Craig et al., 1998). High temporal and spatial resolution in photon detection systems is particularly critical for studies such as transient phenomena in condensed matter (Barty et al., 2008), the radiative decay of molecules (Berezin and Achilefu, 2010), and Time-of-Flight Positron Emission Tomography (TOF-PET) (Lewellen 1998). Recently, we have reported a novel fully digital 3D (x–y–time) imager for hard X-rays (Lusardi et al., 2024). This imager is engineered to achieve improved temporal and spatial resolutions below tens of picoseconds and one hundred micrometers, respectively, by utilizing capacitive coupling of a GaAs separate absorption and multiplication avalanche photodiode (SAM-APD) to cross delay-lines (CDLs). Silicon has traditionally been one of the fundamental materials used in X-ray detectors. However, a major drawback of silicon is that its absorption efficiency drops significantly at photon energies above 15 keV. A promising alternative is GaAs, whose constituent elements have higher atomic numbers than silicon. This results in higher X-ray absorption and enables the use of thinner absorption region which is beneficial for improving temporal resolution. Furthermore, the higher electron mobility in GaAs contributes to reduced response times. However, a known limitation of III–V semiconductors like GaAs is the similar ionization coefficients of electrons and holes, which leads to increased multiplication noise (McIntyre, 1966). To overcome this problem, a GaAs/AlGaAs SAM-APDs were developed (Nichetti et al., 2019). In this device, a multiplying superlattice structure with an energetic staircase profile was designed, which preserves electron multiplication while reducing hole multiplication. In GaAs based devices, the preparation of ohmic contacts is also crucial. Several techniques have been developed to achieve low-resistance, reliable contacts. One common approach involves growth of a highly doped layer (with a concentration above $\sim 1\times10^{18}$ cm$^{-3}$) before depositing the contact material, which provides tunneling effects (Lin et al., 2021). Different materials have been used for different types of doped GaAs to form ohmic contacts. The AuGe/Ni/Au is commonly used to form ohmic contacts on n-type GaAs (Shin et al., 1987). An additional Ni layer between the GaAs substrate and the AuGe has also employed to improve metal adhesion (Murakai, 2002) and to act as a diffusion barrier for Ge and Au (Vidimari, 1979). For p-type GaAs, metal combinations such as Ti/Pt/Au (Lin et al., 2021) and Au/Ni/Au (Jones et al., 2024) have reported as effective materials for achieving ohmic behavior. Cr/Au has also been studied as an ohmic contact for GaAs. When this metal stack was deposited on n-type GaAs and annealed, it exhibited ohmic behavior at 400 °C, which is lower than the annealing temperature required for AuGe/Ni/Au contacts (Guo et al., 2018). For p-type GaAs, Cr/Au contacts were investigated after annealing at 440 °C for 2 minutes. However, the results showed surface roughening and deterioration in the specific contact resistance (Mahajan, et al., 2018). It is well known that roughening of the surface morphology leads to a decrease in the effective area of flow, resulting in an increase in contact resistance values (Baca and Ashby Carol, 2005). U. Schade et al. studied this metal system as well and observed that surface modification begins at approximately 400 °C (Schade et al., 1988). Additionally, the Cr layer remained stable during heat treatments up to 480 °C, but higher temperatures Cr diffused into the Au layer.

In the case of the developed photodetector, due to the particular architecture that will be described later, the challenge was to achieve ohmic contacts on a lightly n-doped and extremely thin GaAs region. The contacts were prepared by depositing Cr/Au, followed by annealing in a nitrogen atmosphere at low temperatures. Thanks to this, it was possible to develop a p+-i-n based photodetector, which is also a preliminary step toward the future SAM APD, already capable of capacitively detecting photon pulses from a high-frequency laser with a temporal jitter on the order of tens of picoseconds. The obtained pulses were designed to match those that will be produced by the multiplication stages upon the arrival of single hard X-ray photons, thereby confirming the validity of the proposed approach.

## 2. Materials and Methods

### 2.1 Capacitively coupled GaAs p+-i-n/substrate photodetector (CC-GaAs PIN/S PD)

*2.1.1 Material growth*

The GaAs PIN (p+-i-n) photodiode structure of the CC-GaAs PIN/S PD was grown by molecular beam epitaxy (MBE). The epitaxial layers of the structure were grown on a semi-insulting GaAs substrate (resistivity of $10^8$ Ω•cm, thickness of 500 μm, and diameter of 2″) at 610 $^0$C. After the growth of a 100 nm thick GaAs buffer layer on the substrate, a 2 μm lightly doped n-type GaAs was grown. Then, a 1 μm intrinsic GaAs was grown. Finally, the structure was completed with a 150 nm highly doped p-type GaAs layer. The detailed layers structure and doping concentrations are summarized in Table 1.

*Table 1: Material, thickness, and doping of the grown GaAs PIN structure*

| Layers | Material | Thickness (nm) | Dopant | Concentration (cm$^{-3}$) |
|---|---|---|---|---|
| p+ layer | GaAs | 150 | C | $6 \times 10^{18}$ |
| i layer | GaAs | 1000 | Undoped | – |
| Lightly doped n-type layer | GaAs | 2000 | Si | $2 \times 10^{16}$ |
| Buffer layer | GaAs | 100 | Undoped | – |

*2.1.2 Fabrication*

The fabricated CC-GaAs PIN PD consisting of a mesa structure of the grown photodiode, ohmic and non ohmic contacts is depicted in figure 1. The mesa with diameter of 1 cm was prepared by optical lithography and etching in $H_3PO_3:H_2O_2:H_2O$. Material deposition was performed using electron-beam evaporation (EBE). The detector fabrication involved two stages of Cr/Au contact deposition. First, 5 nm of chromium (Cr) and 10 nm of gold (Au) were deposited on both the

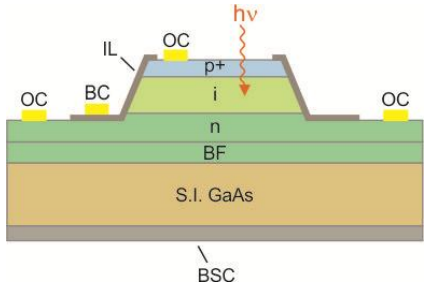

*Figure 1. Capacitively coupled GaAs p+-i-n/substrate photodetector (CC-GaAs PIN/S PD). Buffer (BF), n-type, i-intrinsic, and p-type layers of GaAs are grown on S.I. GaAs substrate. OC-Cr/Au ohmic contacts, BC-Cr/Au blocking contact, IL-isolating $SiO_2$ layer, BSC – backside tin contact. $i_d$ and $i_p$ arrows indicate diode current and pulsed current passing through the detector, respectively.*

p+ and lightly doped n-type layers. To obtain ohmic contacts, a multi-step thermal annealing process was carried out at temperatures below 400 °C in a nitrogen atmosphere. A thermocouple was positioned near the sample to monitor the temperature accurately. The second Cr/Au contact (BC) was deposited following the deposition of an insulating $SiO_2$ layer. This contact is used to suppress shunt currents by applying the same potential as the OC contact on the p$^+$-layer. Additionally, a backside contact (BSC) was formed by depositing tin.

*2.1.3 Operation principle*

To understand the operation of the device, it is useful to refer to the scheme widely employed for photon detection using MicroChannel Plates (MCPs), and in particular to the architecture designed to simultaneously achieve both spatial and temporal resolution (Nichetti, et al., 2019; Cautero et al., 2024). In this architecture, the output pulses from the MCPs, corresponding to the arrival of single particles, reach a resistive anode whose purpose is to collect the current while simultaneously acting as a capacitive coupling element through which the electromagnetic pulses are transmitted to an underlying cross-delay line (CDL). This acquisition scheme fundamentally differs from that of pixel detectors, where each pixel operates independently. In pixel-based detectors, simultaneous detection of multiple photons is possible, and the spatial resolution can, depending on the implementation, reach the sub-micrometer scale. Conversely, detectors based on CDL readout cannot detect multiple photons arriving within the same delay-line propagation time window (typically on the order of a few nanoseconds). However, in many practical applications - particularly those involving non-pulsed or quasi-continuous sources, such as synchrotron radiation facilities where the photon flux can be approximated as continuous - this limitation does not represent a significant drawback. In these scenarios, other aspects become dominant, for which the delay-line-based approach can outperform pixel detectors by orders of magnitude. Among these advantages are the ability to retrieve full x, y, and t information using only four readout channels, and the outstanding timing resolution, which in state-of-the-art delay-line detectors currently reaches values on the order of 10 ps.

To adopt the same approach using a GaAs-based structure instead of MCPs – thereby achieving higher sensitivity to high-energy photons and potential spectroscopic capabilities – the CC-GaAs PIN PD incorporates an n-type region that not only provides reverse bias to the intrinsic layer but is also designed to deplete in a controlled manner, leaving a thin undepleted layer (see fig. 1). This layer maintains the bias across the PIN while simultaneously acting as a resistive layer. Since the PIN structure is deposited on a high-resistivity, semi-insulating GaAs substrate, the overall structure mirrors the resistive anode employed in traditional MCP detector designs with an absorption region. Following the schematic shown in fig. 1, it can be observed that the charges generated by the incoming photons - which, in the forthcoming SAM APD device, will also undergo multiplication - are dissipated through the resistive layer. However, for the transient pulses, the capacitive path through the substrate exhibits a much lower impedance, allowing the signals to be easily detected by anodes (or cross-delay anodes) placed at the back of the insulating substrate.

This detection mechanism can be modelled by its equivalent electronic circuit. The small-signal equivalent circuit of the CC-GaAs PIN PD, with the BSC connected to an oscilloscope, is shown in Figure 2. Under reverse bias, $R_{PD}$ is sufficiently large to be neglected. The resistance $R_n$ increases with the applied reverse bias, as it depends on the depletion width. To ensure proper operation, $R_n$ must be low enough to establish an appropriate DC operating point, yet high enough to allow the photogenerated pulse to propagate through the capacitive path toward the BSC, where it can be detected by the oscilloscope or by anodes (or cross-delay anodes).

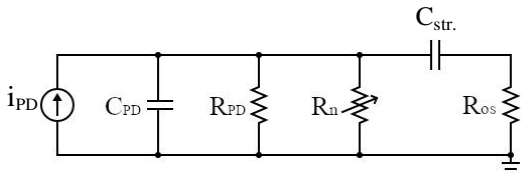

*Figure 2. $i_{PD}$, $C_{PD}$, $R_{PD}$ are photogenerated current, junction capacitance and shunt resistance of the grown PIN photodiode, respectively. $R_n$, $C_{str.}$ and $R_{OS}$ are resistance of n-region, capacitance of substrate and resistance of oscilloscope (50 Ω), respectively.*

The X-ray range that can be detected by the fabricated detector is determined by photon absorption in GaAs. For example, at a photon energy of 20 keV, a 100 μm thick GaAs detector exhibits an absorption greater than 90% (Colja et al., 2022). However, the absorption fraction decreases by approximately 16% at 30 keV (Looker et al., 2019). Therefore, the fabricated detector is expected to be capable of detecting photon energies of up to approximately 30 keV.

## 3. Results

*3.1 Ohmic contact engineering*

The electrical characteristics of the fabricated detector were measured using a B1500A semiconductor device analyzer. In addition to fabricating the CC-GaAs PIN photodiode, we also focused on developing reliable ohmic contacts on lightly doped ($2 \cdot 10^{16}$ cm$^{-3}$) n-type GaAs, which are essential for proper detector operation. Thermal annealing is a crucial step in forming ohmic contacts for GaAs-based devices, as it promotes atomic interdiffusion between the metal layer and the semiconductor, thereby improving contact properties. However, for Cr/Au metallization, annealing temperatures above 400 °C can cause several issues, such as thermal decomposition of GaAs due to arsenic out-diffusion from the surface, surface roughening, and deep metal diffusion toward the n–p interface deteriorating the diode characteristics, etc. Therefore, we focused on multi-step thermal annealing at lower temperatures. At each step, the contacts were annealed for 2 minutes at incrementally higher temperatures until ohmic behavior was achieved. Thus, current–voltage (I–V) measurements were performed in dark condition using two Cr/Au contacts deposited on the n-GaAs layer of the detector (see Figure 1). For comparison, all I–V curves obtained after each annealing step are presented together in Figure 3(A). For a more detailed analysis, these curves are also presented separately. The first curve, obtained before annealing, is shown in Figure 3(B). Its nonlinear behavior is attributed to the formation of a Schottky barrier. The annealing process was started at 280 $^0$C. Although the change in the curve indicates that atomic diffusion occurs even at this temperature, the I–V behavior remains nonlinear. A further increase in temperature to 290 °C does not change this nonlinear behavior (Figure 3(C)). The temperature was then continuously increased until the I–V dependence became linear. As shown in Figure 3(D), the curve becomes linear at 315 °C. With further temperature increase, the change in curve`s slope diminishes, becoming negligible at 330 °C, indicating that the resistance has reached its minimum and the ohmic contact is formed. We also believe that similar results could be achieved by annealing at lower temperatures for longer durations.

## 3.2 Current-voltage and capacitance-voltage characteristics of the PIN structure

The characterization of the PIN structure was carried out using Cr/Au ohmic contacts (OC) deposited on the p$^+$ and n layers (see Figure 1(A)). The diode I–V curve measured in dark condition is shown in Figure 4(A). To minimize shunt currents, the same potential was applied to the blocking contact (BC) as to the ohmic contact on the p$^+$ layer. This approach is expected to suppress leakage current along the surface of the PIN junction. The I–V curves measured with and without BC biasing, are presented in Figure 4(B). Indeed, the current decreases when the BC is biased. This effect becomes particularly noticeable beyond a bias of 4 V. For photoresponse measurements, a laser with a frequency of 80 MHz and a wavelength of 800 nm was used. The I–V characteristic measured under laser irradiation is shown in Figure 5. It also gradually increases with voltage up to a certain threshold. At an intensity of 10 mW, the current saturates beyond 6 mV.

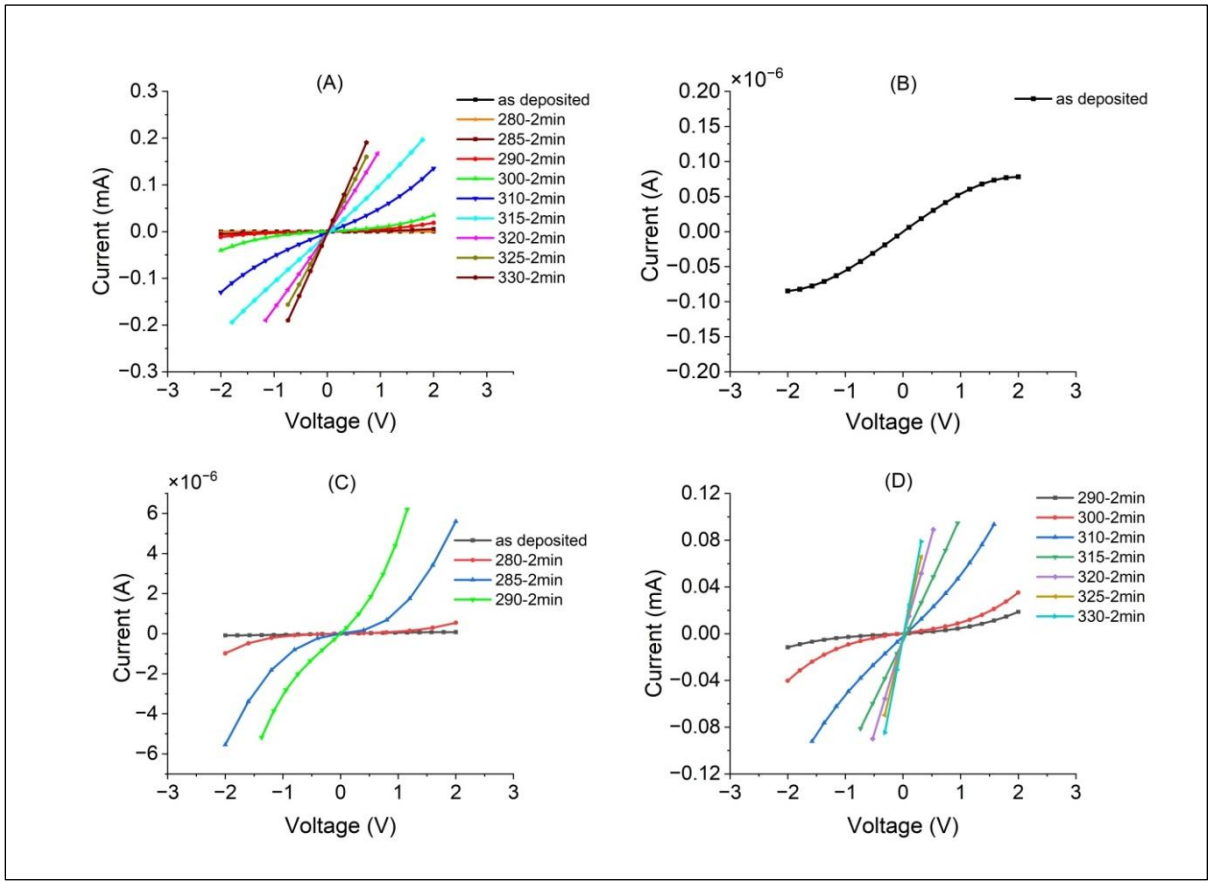

*Figure 3. I-V curves of Cr/Au contact deposited on n-GaAs: (A) – before and after thermal annealing, (B) – before annealing, (C) – before annealing and annealed at 280-290 $^0$C, (D) – annealed at 290-330 $^0$C.*

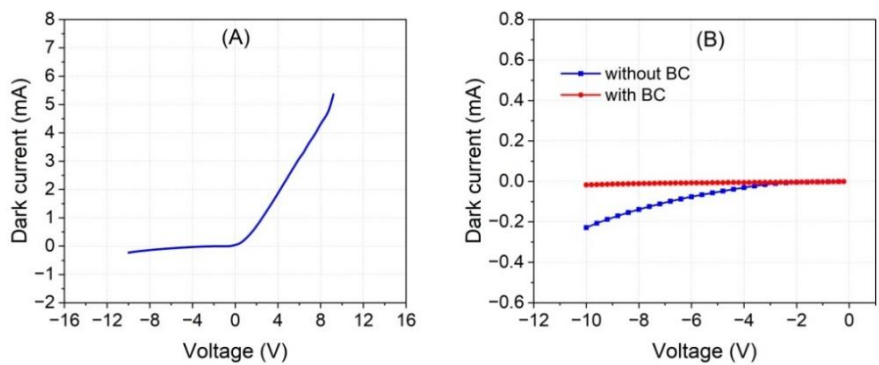

*Figure 4. I-V curves of the GaAs PIN structure: (A) – without BC biasing, (B) – with and without BC biasing.*

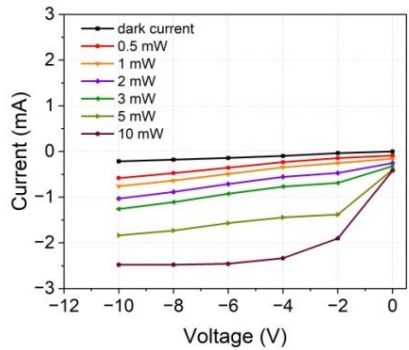

*Figure 5. I-V curves measured under laser irradiation at different intensities.*

Consequently, this reduces both material expense and duration of contact deposition process. Typically, to obtain ohmic contacts on p–n or p–i–n GaAs devices, two separate evaporation processes are carried out to deposit different materials on the n-type and p-type layers, respectively (Lioliou and Barnett, 2016). Moreover, more than two materials are often deposited on at least one of the layers (Shin et al., 1987; Lin et al., 2021). In our case, a single deposition process using only two metals (Cr/Au) was employed for both layers. Consequently, this approach reduces both the material expense and the duration of contact deposition process. The capacitance–voltage (C–V) characteristic measured at 1 MHz in dark condition is shown in Figure 6. The C–V curve indicates that the depletion region widens as the capacitance decreases, up to 1.8 V, beyond which the structure becomes fully depleted. With further increases in bias, the capacitance of the PIN structure remains around 12 pF.

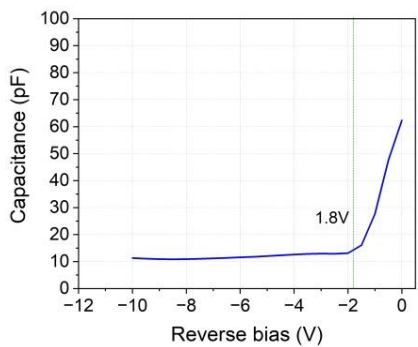

*Figure 6. C-V characteristic of the PIN structure.*

*3.3 Demonstration of operation*

To demonstrate the device operation, the CC-GaAs PIN/S PF was tested under pulsed laser irradiation. The signal pulses detected from the BSC contact (see Figure 1(A)) at different bias voltages and a laser intensity of 10 mW are shown in Figure 7. The pulse amplitude increases with increasing bias, reaching up to 1.25 mV. Given that the laser pulses operated at a frequency of 80 MHz, an output power of 10 mW corresponds to approximately $10^8$ photons per pulse. Taking into account the absorption by the chromium, gold, and $p^+$ GaAs layers, and assuming a "one-to-one" conversion between photons and electrons, this implies that the observed pulses correspond to charge packets of roughly $10^6$ electrons. This value is fully consistent with what can be expected for each hard X-ray photon once a multiplication stage is added and the substrate thickness is reduced. This configuration enables the possibility of position-sensitive detection through additional engineering of the bottom contact. In the forthcoming developments of the project, the backside contact (BSC) will be segmented into multiple contacts to enable irradiation position detection. The CC-GaAs PIN/S PD will be investigated under hard X-ray radiation. The expected results will be used for the fabrication of next-generation hard X-ray imager based on capacitively coupled GaAs SAM-APD and CDLs (Lusardi et al., 2024).

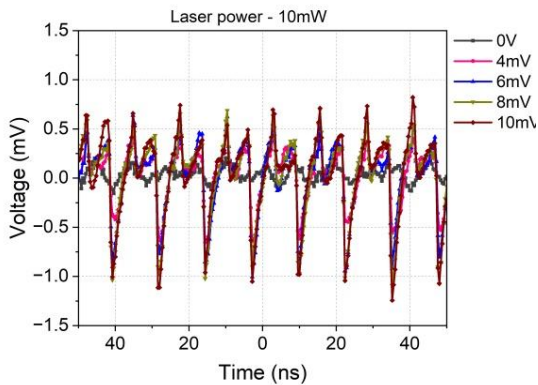

*Figure 7. Voltage pulses obtained from the BSC contact using an oscilloscope terminated at 50 Ω.*

**4. Conclusion**

In this paper, we present a capacitively coupled GaAs p+-i-n/substrate photodetector (CC-GaAs PIN/S PD). This is also the first step toward the development of a GaAs-based SAM-APD device capable of detecting high-energy photons – or, through suitable scintillators, gamma radiation - with high spatial and temporal resolution. The GaAs PIN structure was grown on semi-insulating GaAs substrate by MBE. A multi-step annealing process of the Cr/Au contacts was developed to achieve ohmic behavior on lightly doped n-type GaAs. The annealing was carried out at a low temperature range of 280-330 $^0$C in a nitrogen atmosphere. Such low annealing temperatures allow avoiding deteriorating of the diode characteristics. Simultaneously, the same contact preparation procedure was applied to the p$^+$-GaAs layer. The resulting GaAs PIN structure exhibited typical diode I–V characteristics. Additionally, the CC-GaAs PIN/S PD includes an additional contact that reduces leakage currents by applying the same bias as the anode. Pulses on the order of millivolts were measured in correspondence with charge packets of approximately 10$^6$ electrons, indicating the potential of the CC-GaAs PIN/S PD for high-energy photon detection. This detector is suitable for future applications in hard X-ray imaging.

**Acknowledgments**

This work was funded by Italian MIUR through PRIN 20227N9LW7, and the EU through Next Generation EU, Mission 4, component 1, CUP B53D23002300006, Linea di investimento 1.3, Progetto PE00000023_NQSTI "National Quantum Science & Technology Institute" and M4C2, within the PNRR project NFFA-DI, CUP B53C22004310006, IR0000015. Also, this work was supported by the Science Committee of RA (Research project № 23PTS-2F007).